\documentclass[a4paper,preprint,onecolumn,aps,pra,superscriptaddress]{revtex4-1}
\usepackage{xcolor}

\usepackage[english]{babel}
\usepackage[T1]{fontenc}

\usepackage{amsmath, amssymb, amsthm, mathtools}
\usepackage{bm}
\usepackage{braket}
\usepackage{siunitx}

\usepackage{graphicx}
\usepackage{csquotes}
\usepackage{xcolor}
\usepackage[colorlinks=true, linkcolor=blue, citecolor=blue, urlcolor=blue]{hyperref}
\usepackage{subcaption}

\begin{document}

\raggedbottom

\definecolor{pascal}{RGB}{204,121,167}
\newcommand{\PS}[1]{\textcolor{pascal}{#1}}

\definecolor{alec}{RGB}{0, 200, 132}
\newcommand{\AO}[1]{\textcolor{alec}{#1}}

\definecolor{etienne}{RGB}{150, 0, 250}
\newcommand{\EG}[1]{\textcolor{etienne}{[EG]#1}}

\title{Large-scale NMR simulation on a trapped-ion quantum computer}

\author{Pascal Stadler}
    \thanks{These authors contributed equally to this work.}
    \affiliation{HQS Quantum Simulations GmbH, Rintheimer Stra{\ss}e 23, 76131 Karlsruhe, Germany}

\author{Alec Owens}
    \thanks{These authors contributed equally to this work.}
    \email{alec.owens@quantinuum.com}
    \affiliation{Quantinuum, Terrington House, 13-15 Hills Road, Cambridge CB2 1NL, UK}

\author{Etienne Granet}
    \affiliation{Quantinuum, Leopoldstra{\ss}e 180, 80804 Munich, Germany}

\author{David Mu\~{n}oz Ramo}
    \affiliation{Quantinuum, Terrington House, 13-15 Hills Road, Cambridge CB2 1NL, UK}

\author{Michael Marthaler}
    \affiliation{HQS Quantum Simulations GmbH, Rintheimer Stra{\ss}e 23, 76131 Karlsruhe, Germany}

\date{\today} 

\begin{abstract}
Simulating nuclear magnetic resonance (NMR) spectra is a promising application of quantum simulation. Using Quantinuum's System Model H2, a trapped-ion quantum computer, we demonstrate an end-to-end, large-scale digital NMR simulation of a classically challenging benchmark molecule, 1,2-di-\emph{tert}-butyl-diphosphane. We implement a hardware-efficient reduction of the nuclear-spin Hamiltonian, enabling Trotterized real-time evolution of an effective 21-spin model with tailored error suppression to reduce the effects of device noise. The reconstructed liquid-state proton NMR spectrum agrees with classical reference calculations and reproduces key spectroscopic features that previous quantum hardware demonstrations did not capture. Given the widespread use of NMR in chemical analysis and industrial research, these results advance digital quantum simulation of NMR spectra towards practical quantum utility.
\end{abstract}

\maketitle

\section{Introduction}

Quantum computers provide a natural platform for simulating the dynamics of interacting spin systems, directly representing quantum states and their evolution without the exponential resource scaling that limits classical approaches. The potential for quantum speed-up in such simulations is widely anticipated~\cite{Childs2018}. This makes nuclear magnetic resonance (NMR) spectroscopy a compelling application for quantum simulation~\cite{hqs_usecase}, since it probes coupled nuclear-spin dynamics to yield detailed insight into local atomic structure and chemical environments~\cite{Das2025}. NMR is an essential analytical tool in chemistry and materials science, with broad industrial relevance in areas such as drug discovery~\cite{Pellecchia2002} and battery materials research~\cite{Pecher2017}.

Interest in the quantum simulation of NMR experiments is growing~\cite{hqs_usecase,hqs_nmr,Castillo2024,Khedri2024,Seetharam2023,Burov2024,Burov2025,Burov2026,fujitsu_nmr,PRXQuantum.3.030345,google_otoc,Elenewski2026,AppQsim}. There have been demonstrations on quantum hardware~\cite{Khedri2024,Seetharam2023,Burov2024,Burov2025,Burov2026}, novel applications of quantum algorithms and quantum-enabled methods to NMR workflows~\cite{fujitsu_nmr,PRXQuantum.3.030345,google_otoc}, comprehensive resource estimates for NMR spectroscopy in the fault-tolerant era~\cite{Elenewski2026}, and the use of NMR simulation as an application-oriented benchmark test of quantum computers~\cite{AppQsim}. However, the specific regimes in which quantum computers can provide a practical advantage for NMR simulation remain under debate. These include zero- to ultralow-field (ZULF) settings, where internal spin–spin couplings dominate, and solid-state environments governed by dipolar spin interactions, both of which can lead to complex dynamics that are difficult to simulate classically.

High-field liquid-state NMR spectroscopy, which is the focus of this work, is among the most widely used and experimentally accessible techniques. In this regime, classical simulation methods are highly optimized, exploiting physically motivated approximations to reduce the computational cost to linear~\cite{Castillo2011,hqs_nmr} or polynomial~\cite{spinach2011} scaling with the number of nuclear spins. Nevertheless, these efficiencies rely on assumptions that can break down in the presence of strongly interacting systems, with the underlying Hilbert space still growing exponentially with system size. High-field liquid-state NMR can therefore serve as a stringent benchmark for quantum simulation methods, combining chemically interesting spectra with well-established classical references.

A known classically challenging molecule to simulate is 1,2-di-\emph{tert}-butyl-diphosphane, referred to here as diphosphane, a heteronuclear \(22\)-spin system with unique characteristics, discussed below. Diphosphane's NMR spectrum was recently simulated in a 22-qubit experiment with error mitigation and suppression on IBM superconducting and IonQ trapped-ion quantum computers~\cite{Burov2025}, representing an important demonstration of large-scale NMR simulation on quantum hardware. The resulting spectra, however, displayed noticeable discrepancies compared to classical references. In particular, the proton NMR spectrum failed to reproduce the $3.9$--$4.3$~ppm (parts per million) region, which exhibits a weaker but distinctive double-peak structure. This structure encodes key information about the molecule, making its accurate reproduction a central motivation for the present work. Although diphosphane's spectrum can be simulated by efficient classical methods with approximations~\cite{spinach2011,hqs_nmr}, it is significantly more demanding than some of the small textbook spin systems used in previous quantum hardware experiments, such as the four-qubit simulations of acetonitrile~\cite{Seetharam2023}.

In this work, we use Quantinuum H2-1, a trapped-ion quantum computer~\cite{PhysRevX.13.041052}, to simulate the proton NMR spectrum of diphosphane. We implement Trotterized real-time evolution of a hardware-efficient 21-spin effective Hamiltonian in an experiment using 21 system qubits and 21 ancilla qubits. The simulations employ up to 70 Trotter steps, with the largest circuits reaching a two-qubit gate depth of 1442. These are among the deepest circuits yet executed for an NMR simulation on quantum hardware, and tailored error suppression is essential for preserving signal fidelity over extended evolution times. This enables an end-to-end workflow in which time-domain correlation functions are generated on the quantum device and processed using standard Fourier-based NMR analysis. To our knowledge, these are the most accurate large-scale NMR simulation results reported on a quantum computer to date, highlighting the potential of quantum hardware for NMR simulation.

\section{Model and methodology}

\subsection{Diphosphane as a benchmark for quantum NMR simulation}

\begin{figure}[h]
    \centering
    \includegraphics[width=0.5\columnwidth]{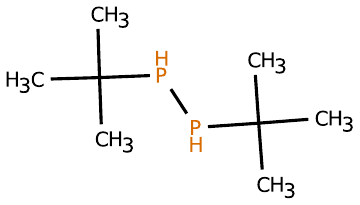}
    \caption{Molecular structure of 1,2-di-\emph{tert}-butyl-diphosphane, used in this work as a benchmark heteronuclear spin system for NMR simulation. The molecule contains two coupled \(^{31}\)P nuclei and twenty \(^{1}\)H nuclei. The phosphorus atoms are highlighted to emphasize the two-spin subsystem that is later transformed to the singlet--triplet basis and projected onto an effective degree of freedom.}
    \label{fig:diphosphane_structure}
\end{figure}

Diphosphane, illustrated in Fig.~\ref{fig:diphosphane_structure}, is a heteronuclear \(22\)-spin molecule containing two \(^{31}\)P and twenty \(^{1}\)H nuclei with spin quantum number $I=1/2$. It is a notable test case to probe the practical limits of both classical and quantum workflows for three reasons. First, as a heteronuclear system, its Hamiltonian contains clearly separated energy scales associated with the phosphorus and proton Zeeman terms, i.e., the single-spin interactions with the external magnetic field. This is the regime in which Trotterized real-time evolution becomes challenging as the time step must be small enough to resolve fast dynamics, while the total evolution time must be long enough to resolve slow spectral features. Second, the molecule exhibits substantial symmetry and contains multiple magnetically equivalent protons, which enables physically motivated model reduction without eliminating the dominant spectral signatures. Third, the proton subsystem is not fully connected and important correlations are mediated through the two phosphorus centers.

\subsection{Spin Hamiltonian of diphosphane}

We partition the spins of diphosphane into two phosphorus nuclei \(P_A\) and \(P_B\), two protons \(h_A\) and \(h_B\) bonded to the phosphorus nuclei, and two sets of nine equivalent methyl protons,
\[
\mathcal{M}_A=\{m^A_{1},\dots,m^A_{9}\},
\qquad
\mathcal{M}_B=\{m^B_{1},\dots,m^B_{9}\},
\]
where $m$ labels the protons in each set. The Zeeman part of the Hamiltonian is determined by three chemical shifts: \(\delta_P\) for the two \(^{31}\)P nuclei, \(\delta_h\) for the two protons \(h_A\) and \(h_B\) connected to the P atoms, and \(\delta_m\) for the methyl protons. The scalar-coupling network is described by four coupling constants: the phosphorus--phosphorus coupling \(J_{PP}\), the direct phosphorus--proton coupling \(J_{\mathrm{dir}}\), the weaker cross coupling \(J_{\mathrm{cross}}\), and the phosphorus--methyl coupling \(J_m\). These parameters were taken from the {\it Spinach} software library~\cite{spinach2011} and are listed in Table~\ref{tab:diphosphane_parameters}. Chemical shifts \(\delta\) in ppm are converted to frequency according to \(\Delta \nu = \delta\,\nu_{\mathrm{ref}}\times 10^{-6}\) in Hz, or equivalently \(\Delta \omega = 2\pi \Delta \nu\) in rad\,s\(^{-1}\), where \(\nu_{\mathrm{ref}}\) is the reference Larmor frequency of the  nucleus.

\begin{table}[h]
\caption{Parameters of the diphosphane spin Hamiltonian of Eq.~\eqref{eq:H_dip_compact}.}
\label{tab:diphosphane_parameters}
\begin{ruledtabular}
\begin{tabular}{@{} c c r @{}}
\textbf{Symbol} & \textbf{Assignment} & \multicolumn{1}{c}{\textbf{Value}} \\
\hline
$\delta_P$ & $^{31}$P chemical shift & $-43.844~\mathrm{ppm}$ \\
$\delta_h$ & Phosphorus-proton chemical shift & $4.09~\mathrm{ppm}$ \\
$\delta_m$ & Methyl-proton chemical shift & $1.354~\mathrm{ppm}$ \\
$J_{PP}$ & $P_A$--$P_B$ coupling & $301.99~\mathrm{Hz}$ \\
$J_{\mathrm{dir}}$ & Direct P--H coupling & $-321.62~\mathrm{Hz}$ \\
$J_{\mathrm{cross}}$ & Cross P--H coupling & $-19.15~\mathrm{Hz}$ \\
$J_m$ & P--methyl coupling & $15.63~\mathrm{Hz}$ \\
\end{tabular}
\end{ruledtabular}
\end{table}

The chemical shifts in Table~\ref{tab:diphosphane_parameters} determine the Zeeman angular frequencies entering the Hamiltonian. For a spin \(k\) of isotope \(\alpha(k)\), we use
\begin{equation}
\omega_k
=
\gamma_{\alpha} B_0(1-\sigma_k)
\simeq
\omega_{\alpha,\mathrm{ref}}
\left(1+\delta_k\times10^{-6}\right),
\label{eq:chemical_shift_frequency}
\end{equation}
where \(B_0\) is the static magnetic field, \(\gamma_{\alpha}\) is the gyromagnetic ratio, \(\sigma_k\) is the shielding constant, and \(\delta_k\) is the chemical shift in ppm. Here, \(\omega_{\alpha,\mathrm{ref}}=2\pi\nu_{\alpha,\mathrm{ref}}\) is the angular reference frequency for isotope \(\alpha\). Our simulations use a \(500~\mathrm{MHz}\) \(^{1}\mathrm{H}\) spectrometer field, corresponding to \(B_0\simeq11.74~\mathrm{T}\). 

The diphosphane Hamiltonian, expressed in angular-frequency units, is
\begin{align}
\hat H_{\mathrm{dip}}
&=
-\omega_P\!\left(\hat I_{P_A}^z+\hat I_{P_B}^z\right)
-\omega_h\!\left(\hat I_{h_A}^z+\hat I_{h_B}^z\right)
-\omega_m\!\sum_{j\in \mathcal{M}_A\cup\mathcal{M}_B}\hat I_j^z
\nonumber\\
&\quad
+2\pi J_{PP}\,\hat{\mathbf I}_{P_A}\!\cdot\!\hat{\mathbf I}_{P_B}
+2\pi J_{\mathrm{dir}}
\left(
\hat{\mathbf I}_{P_A}\!\cdot\!\hat{\mathbf I}_{h_A}
+
\hat{\mathbf I}_{P_B}\!\cdot\!\hat{\mathbf I}_{h_B}
\right)
\nonumber\\
&\quad
+2\pi J_{\mathrm{cross}}
\left(
\hat{\mathbf I}_{P_A}\!\cdot\!\hat{\mathbf I}_{h_B}
+
\hat{\mathbf I}_{P_B}\!\cdot\!\hat{\mathbf I}_{h_A}
\right)
+2\pi J_m \sum_{j\in\mathcal{M}_A}\hat{\mathbf I}_{P_A}\!\cdot\!\hat{\mathbf I}_j
+2\pi J_m \sum_{j\in\mathcal{M}_B}\hat{\mathbf I}_{P_B}\!\cdot\!\hat{\mathbf I}_j .
\label{eq:H_dip_compact}
\end{align}
Here \(\hat{\mathbf I}_k=(\hat I_k^x,\hat I_k^y,\hat I_k^z)\) denotes the spin-\(\tfrac12\) operator of nucleus \(k\). 

For the quantum-computing implementation, each spin-\(\tfrac12\) operator is represented by Pauli operators according to
\(
\hat I_k^\alpha = \frac{1}{2}\hat\sigma_k^\alpha,
\)
where
\(
\alpha\in\{x,y,z\},
\)
and \(\hat\sigma_k^\alpha\in\{\hat X_k,\hat Y_k,\hat Z_k\}\) acts on the qubit representing nuclear spin \(k\). 
The quantities \(\omega_P\), \(\omega_h\), and \(\omega_m\) are the chemical-shift-dependent angular Zeeman frequencies of the phosphorus, the P--H proton, and methyl-proton spins, respectively, as defined in Eq.~\eqref{eq:chemical_shift_frequency}. The physical structure of the problem is made explicit by Eq.~\eqref{eq:H_dip_compact}, highlighting the various interactions between the nuclei in diphosphane. A prominent interaction scale is the strong \(^{31}\mathrm{P}\)--\(^{31}\mathrm{P}\) scalar coupling, which is comparable in magnitude to the direct \(^{31}\mathrm{P}\)--\(^{1}\mathrm{H}\) coupling; each phosphorus nucleus is strongly coupled to one proton, weakly coupled to the proton on the opposite side of the molecule, and coupled with equal strength to the nine methyl protons of its own \emph{tert}-butyl group.

\subsection{Hardware-efficient projection of the phosphorus pair}

A direct hardware implementation of the full \(22\)-spin Hamiltonian is costly because the heteronuclear separation of energy scales leads to deep Trotterized time evolution circuits. To obtain a model that is more suitable for present-day hardware, we rewrite the two-phosphorus subsystem in the singlet--triplet basis~\cite{Levitt}
\[
\bigl\{
\ket{T_+},\;
\ket{T_0},\;
\ket{T_-},\;
\ket{S}
\bigr\},
\]
where the basis states are given by
\begin{equation}
\begin{aligned}
\ket{T_+} &= \ket{\uparrow\uparrow}, \\
\ket{T_0} &= \frac{\ket{\uparrow\downarrow}+\ket{\downarrow\uparrow}}{\sqrt{2}}, \\
\ket{T_-} &= \ket{\downarrow\downarrow}, \\
\ket{S}   &= \frac{\ket{\uparrow\downarrow}-\ket{\downarrow\uparrow}}{\sqrt{2}} .
\end{aligned}
\end{equation}

Let \(\hat U_{\mathrm{ST}}\) denote the corresponding Schur transform acting on the phosphorus pair. The Hamiltonian in the singlet--triplet basis is
\begin{equation}
\hat H_{\mathrm{ST}}
=
\hat U_{\mathrm{ST}}\,
\hat H_{\mathrm{dip}}\,
\hat U_{\mathrm{ST}}^\dagger.
\end{equation}
We project onto the two-dimensional subspace spanned by \(\ket{S}\) and \(\ket{T_0}\), i.e., ${\rm span}(\ket{01} + \ket{10},\ket{01} - \ket{10} )$, and define the effective qubit states
\[
\ket{0_{\mathrm{eff}}} \equiv \ket{S},
\qquad
\ket{1_{\mathrm{eff}}} \equiv \ket{T_0}.
\]
The corresponding projector is
\begin{equation}
\hat P_{ST_0}
=
\ket{0_{\mathrm{eff}}}\bra{0_{\mathrm{eff}}}
+
\ket{1_{\mathrm{eff}}}\bra{1_{\mathrm{eff}}} ,
\end{equation}
and the projected Hamiltonian is given by
\begin{equation}
\hat H_{\mathrm{proj}}
=
\hat P_{ST_0}\,
\hat H_{\mathrm{ST}}\,
\hat P_{ST_0}^\dagger .
\label{eq:H_proj_def}
\end{equation}

Importantly, this construction can be carried out directly in the operator representation of the Hamiltonian. One applies the Schur transformation and the projection locally to the two-spin phosphorus sector, and then re-expresses the resulting operator in the Pauli basis of the reduced system. In this way, the reduction is performed at the operator level without explicitly constructing the full Hilbert space. The procedure removes the \(\ket{T_+}\) and \(\ket{T_-}\) components and replaces the original two-spin phosphorus subsystem with a single qubit. The spin count is thereby reduced from \(22\) physical spins to \(21\) effective spins. 

In practice, this reduction significantly lowers the gate cost of the Trotterized time evolution circuits while preserving the relevant spectral features of diphosphane. For an all-to-all connected architecture like Quantinuum H2-1 with native parameterized angle \(\mathsf{ZZPhase}(\alpha)\) two-qubit gates, the reduced Hamiltonian requires \(20\) two-qubit gates per Trotter step, compared with \(69\) two-qubit gates for the full unreduced model. The gate count can be understood from the Hamiltonian structure. The full model contains 23 isotropic Heisenberg couplings giving \(69\) entangling operations per Trotter step. The projected model instead contains 20 effective \(\hat{X}_0\hat{Z}_j\) couplings, see Eq.~\eqref{eq:H_proj_compact}, each compiled using a single entangling gate resulting in 20 two-qubit gates per Trotter step. On architectures with restricted connectivity such as superconducting devices, the number of two-qubit gates would increase further due to additional SWAP operations. 

The projection is a hardware-motivated approximation but adds only marginal deviations from the full result in the present parameter regime. Its purpose is to retain the dominant proton resonances in the spectral regions of interest at a circuit depth feasible on current hardware. In a simplified picture, one \(^{1}\mathrm H\) nucleus coupled to the two \(^{31}\mathrm P\) nuclei gives spectral contributions conditioned on the four phosphorus configurations \(\ket{\uparrow\uparrow}\), \(\ket{\uparrow\downarrow}\), \(\ket{\downarrow\uparrow}\), and \(\ket{\downarrow\downarrow}\). The \(ST_0\) projection retains the antiparallel sector, \(\ket{S}\) and \(\ket{T_0}\), but removes the parallel sectors \(\ket{T_+}=\ket{\uparrow\uparrow}\) and \(\ket{T_-}=\ket{\downarrow\downarrow}\). Peaks associated with the discarded sectors are therefore absent in the projected model. However, these discarded parallel sectors are comparatively simple: in a first-order picture, the phosphorus spins are fully polarized and act only as static \(z\)-field shifts on the protons, so their corresponding peak positions can be estimated analytically. Comparison with the unreduced model confirms that the main benchmark peaks are preserved. A similar hardware-motivated reduction strategy was also successfully applied to the simulation of spin-wave spectra in magnetic materials on quantum hardware~\cite{Stadler2026}.

\subsection{Projected Hamiltonian}

For the qubit ordering returned by our implementation, the projected Hamiltonian can be written compactly in Pauli form as
\begin{widetext}
\begin{equation}
\hat H_{\mathrm{proj}}
=
h_0 \hat Z_0
+h_h\left(\hat Z_1+\hat Z_2\right)
+h_m\sum_{j=3}^{20}\hat Z_j
+J_h\,\hat X_0\left(\hat Z_1-\hat Z_2\right)
+J_m^{\mathrm{eff}}\,\hat X_0\sum_{j=3}^{20}(-1)^j \hat Z_j ,
\label{eq:H_proj_compact}
\end{equation}
\end{widetext}
where \(\hat X_j\) and \(\hat Z_j\) denote Pauli operators acting on qubit \(j\). The numerical coefficients of Eq.~\eqref{eq:H_proj_compact} are listed in Table~\ref{tab:projected_coefficients}. All values are reported in the same units (rad{\,}s$^{-1}$) as the Pauli decomposition of the Hamiltonian. The original phosphorus-proton couplings are isotropic and therefore of Heisenberg form, \(\hat{\mathbf I}_{P}\cdot\hat{\mathbf I}_{H}\). After projection onto the \(\{\ket{S},\ket{T_0}\}\)  subspace of the phosphorus pair, the transverse phosphorus operators satisfy \(\hat P_{ST_0}\hat I_{P}^{x}\hat P_{ST_0}
=
\hat P_{ST_0}\hat I_{P}^{y}\hat P_{ST_0}=0\), so the transverse contributions are projected out. Only the \(z\)-component remains, which becomes an \(\hat X_0 \hat Z_j\) coupling. 

\begin{table}[h]
\caption{Coefficients of the projected Hamiltonian in Eq.~\eqref{eq:H_proj_compact}. All values are in angular-frequency units (\(\mathrm{rad}\,\mathrm{s}^{-1}\)). The local field coefficients \(h_h\) and \(h_m\) correspond directly to the proton chemical shifts \(\delta_h\) and \(\delta_m\) listed in Table~\ref{tab:diphosphane_parameters}, related via \(h_\alpha = -\gamma_H(1+\delta_\alpha)B_0/2\), where \(B_0\) is the static external magnetic field and \(\gamma_H\) is the gyromagnetic ratio of \(^{1}\mathrm H\). For the present \(500~\mathrm{MHz}\) proton NMR, \(B_0 = 11.7433~\mathrm{T}\).}
\label{tab:projected_coefficients}
\begin{ruledtabular}
\begin{tabular}{@{} c c r @{}}
\textbf{Symbol} & \textbf{Associated term} & \multicolumn{1}{c}{\textbf{Value}} \\
\hline
$h_0$ & $\hat Z_0$ & $-948.73$ \\
$h_h$ & $\hat Z_1+\hat Z_2$ & $-6424.56$ \\
$h_m$ & $\sum_{j=3}^{20}\hat Z_j$ & $-2126.86$ \\
$J_h$ & $\hat X_0(\hat Z_1-\hat Z_2)$ & $475.12$ \\
$J_m^{\mathrm{eff}}$ & $\hat X_0\sum_{j=3}^{20}(-1)^j\hat Z_j$ & $24.55$ \\
\end{tabular}
\end{ruledtabular}
\end{table}

The Hamiltonian in Eq.~\eqref{eq:H_proj_compact} is the form used to generate the quantum circuits. After projection, the phosphorus pair is represented by a single qubit coupled to two distinct proton qubits and to a set of symmetry-related proton modes.

\subsection{Time-domain NMR correlation functions on a quantum computer}

A detailed derivation of the NMR signal for a pulse-acquire experiment and its formulation in terms of correlation functions can be found in Ref.~\cite{hqs_nmr}. Here we describe only the form implemented in our quantum-computing workflow.

We consider a static magnetic field along the \(z\)-direction and define the collective spin operators
\begin{equation}
\hat I_\alpha = \sum_k \hat I_k^\alpha,
\qquad \alpha\in\{x,y,z\}.
\end{equation}
At thermal equilibrium, the nuclear-spin system density matrix is
\(
\hat\rho_\beta
=
e^{-\beta \hbar \hat H}/Z
\)
with
\(
Z=\operatorname{Tr}\!\left[e^{-\beta \hbar \hat H}\right]
\)
and
\(\beta = 1/k_b T\), where \(k_b\) is the Boltzmann constant and \(T\) is the temperature.
In the high-temperature limit relevant for liquid-state NMR, the density matrix can be expanded as
\begin{equation}
\hat\rho_\beta
\approx
\frac{1}{Z}\left(1-\beta \hbar \hat H\right) \, .
\end{equation}
Since the Zeeman contribution dominates the thermal state, and because we consider a proton NMR experiment, this reduces to
\(
\hat\rho_\beta \propto \hat{1} + \epsilon \hat I_z,
\)
where \(\hat I_z\) denotes the proton contribution to the longitudinal magnetization and \(\epsilon \ll 1\). The identity contribution does not affect the measured signal and will therefore be omitted in the following.

The radiofrequency pulse is represented by a unitary rotation acting only on the proton spins,
\(
\hat R_y\left(\frac{\pi}{2}\right) =e^{-i\frac{\pi}{2} \hat I_y}.
\)
We define the initial density matrix for the subsequent free evolution as the state immediately after the pulse,
\begin{equation}
\hat\rho(0)
=
\hat R_y\!\left(\frac{\pi}{2}\right)
\hat\rho_\beta
\hat R_y^\dagger\!\left(\frac{\pi}{2}\right).
\end{equation}
Omitting the identity contribution, this gives
\begin{equation}
\hat\rho(0)
\propto
\hat R_y\!\left(\frac{\pi}{2}\right)
\hat I_z
\hat R_y^\dagger\!\left(\frac{\pi}{2}\right)
\propto
\hat I_x.
\label{eq:rho0_after_pulse}
\end{equation}

During the subsequent time evolution, this transverse magnetization develops components along both transverse directions. For this reason, the full free-induction decay (FID) must be reconstructed from two correlation functions, corresponding to the \(x\)- and \(y\)-quadratures of the signal.

After the pulse, the deviation density matrix is proportional to \(\hat I_x\), cf.~Eq.~\eqref{eq:rho0_after_pulse}. The thermal factor only sets the overall signal amplitude and is omitted in the following.
The time-domain signal is reconstructed from the following correlation functions~\cite{hqs_nmr}
\begin{equation}
C_{xx}(t)=\mathrm{Tr}\!\left[\hat I_x(t)\hat I_x\right],
\qquad
C_{yx}(t)=\mathrm{Tr}\!\left[\hat I_y(t)\hat I_x\right],
\label{eq:corr_xx_yx}
\end{equation}
where
\begin{equation}
\hat I_\alpha(t)=e^{+i\hat H_{\mathrm{proj}} t}\hat I_\alpha e^{-i\hat H_{\mathrm{proj}} t},
\end{equation}
denotes the Heisenberg-evolved observable, and \(\hat H_{\mathrm{proj}}\) is the projected Hamiltonian used in the simulation. The complex FID is reconstructed as
\begin{equation}
s(t)\propto C_{xx}(t)+i\,C_{yx}(t),
\label{eq:fid_reconstructed}
\end{equation}
and the corresponding NMR spectrum is obtained from the Fourier transform of \(s(t)\).

For the quantum-computing implementation, we expand the trace in an eigenbasis of \(\hat I_x\). Let \(\{\ket{\lambda_i^{(x)}}\}\) satisfy
\begin{equation}
\hat I_x \ket{\lambda_i^{(x)}} = \lambda_i^{(x)} \ket{\lambda_i^{(x)}} .
\label{eq:ix_eigenbasis}
\end{equation}
Then, for \(\alpha\in\{x,y\}\),
\begin{equation}
C_{\alpha x}(t)
=
\sum_i \lambda_i^{(x)}
\bra{\lambda_i^{(x)}}\hat I_\alpha(t)\ket{\lambda_i^{(x)}} .
\label{eq:corr_sampling_x}
\end{equation}
This is the form evaluated on the quantum computer.

\subsection{Trotterization and choice of initial states}

The simulations proceeded as follows. An initial state in the \(\hat I_x\) eigenbasis is prepared. The state is evolved for a time \(t\) under the Trotterized Hamiltonian \(\hat H_{\mathrm{proj}}\), after which the transverse \(x\)- and \(y\)-quadratures are measured. Repeating this procedure for different evolution times, and different initial states, yields estimates of \(C_{xx}(t)\) and \(C_{yx}(t)\), which are combined according to Eq.~\eqref{eq:fid_reconstructed} to reconstruct the full time-domain signal.

In a standard proton NMR experiment, the excitation pulse acts primarily on the \(^{1}\)H nuclei, while the \(^{31}\)P nuclei are not directly excited. Therefore, we restrict the initial-state preparation to the \(^{1}\)H nuclei and keep the degrees of freedom related to phosphorus fixed. The expression in Eq.~\eqref{eq:corr_sampling_x} still contains exponentially many initial states, so evaluating the full sum is not possible. Exploiting the symmetry of the Hamiltonian we restrict the sampling to states with positive transverse magnetization and approximate the sum by a small random subset. Because of the high symmetry of diphosphane and the presence of many magnetically equivalent protons, sampling 10 initial states yields a stable spectrum.

For the time evolution, we utilized a first-order Lie-Trotter decomposition of the projected Hamiltonian~\cite{Trotter1959, Suzuki1976}. The Trotter step must be chosen sufficiently small so that the error arising from the non-commutativity of the Hamiltonian terms remains controlled. In practice, the Trotter step was chosen empirically by two competing requirements: (1) \(\Delta t\) should be as large as possible to reach long evolution times, (2) \(\Delta t\) must remain small enough so that the Trotter error does not distort the spectrum. For the projected Hamiltonian used here, we employed a Trotter step size of \(\Delta t = 0.42~\mathrm{ms}\) based on the agreement between the resulting spectrum and noiseless reference simulations.

For the time evolution, we utilized a first-order Lie--Trotter decomposition of the projected Hamiltonian~\cite{Trotter1959,Suzuki1976}. Expressing the Hamiltonian as a sum of generally non-commuting terms,
\(
\hat H_{\mathrm{proj}} = \sum_\ell \hat H_\ell,
\)
a single Trotter step of duration \(\Delta t\) is approximated as
\begin{equation}
e^{-i \hat H_{\mathrm{proj}} \Delta t}
\approx
\prod_\ell e^{-i \hat H_\ell \Delta t},
\end{equation}
where \(\Delta t\) denotes the discrete time step used in the Trotterized time-evolution.

\section{Hardware experiment}

All circuits were optimized and compiled into Quantinuum's native gateset using the \texttt{pytket} compiler~\cite{tket2020}. Circuits were entirely composed of single-qubit rotation gates,
\begin{align*}
    {\sf Rz}(\alpha) &= \exp(-\frac{1}{2}i\pi\alpha\hat{Z}),\\
    {\sf PhasedX}(\alpha,\beta) &= {\sf Rz}(\beta){\sf Rx}(\alpha){\sf Rz}(-\beta),
\end{align*}
together with the parameterized angle two-qubit entangling gate,
\begin{equation*}
    {\sf ZZPhase}(\alpha)=\exp(-\frac{1}{2}i\pi\alpha\hat{Z}\otimes\hat{Z}).
\end{equation*}
Here, ${\sf Rx}(\alpha)=\exp(-\frac{1}{2}i\pi\alpha\hat{X})$, $\hat{X}$ and $\hat{Z}$ are the standard Pauli operators, and $\alpha$ and $\beta$ are rotation angles.

Prior to the hardware experiment, the number of Trotter steps $N_{\rm t}$ and the number of measurement shots per circuit $N_{\rm shot}$ were optimized under constraints imposed by hardware resources and cost. Since the Trotter timestep $\Delta t$ is fixed, increasing $N_{\rm t}$ extends the maximum sampled duration of the FID as each Trotter step is another sampled time point of the FID, at the expense of greater circuit depth. Increasing $N_{\rm shot}$ reduces statistical (shot) noise in the measurement of the FID. These parameters are related as the total shot budget for the experiment scales with $N_{\rm t} N_{\rm shot}$.

Several parameter pairs $(N_{\rm t}, N_{\rm shot})$ with $37\leq N_{\rm t}\leq 82$ and $20\leq N_{\rm shot}\leq 100$ were evaluated using the end-to-end NMR simulation pipeline in conjunction with the Quantinuum H2-1E emulator, which employs an accurate noise model of the H2-1 device. Generated spectra were compared against a noiseless reference spectrum computed with $N_{\rm t}=100$ and $N_{\rm shot}=1000$. This reference spectrum was not fully converged and served only as a stable benchmark computed with resources beyond the hardware budget to guide parameter selection. Analysis indicated that it was preferable to reduce $N_{\rm shot}$ and increase $N_{\rm t}$ in order to sample more time points and improve the agreement with the noiseless reference, as the simulated NMR spectrum was relatively robust to shot noise, see Fig.~\ref{fig:spectra_shot_noise} in the Appendix. The final parameters chosen for the hardware experiment were $N_{\rm t}=70$ Trotter steps and $N_{\rm shot}=25$ shots per circuit, which provided margin for shot post-selection from leakage errors during hardware execution, discussed below.

\subsection{Error suppression techniques}

Exploratory NMR simulations using the Quantinuum H2-1E emulator revealed that noise induced shifts in the spectroscopic peaks relative to the noiseless reference spectrum, with memory error identified as the dominant error source, see Fig.~\ref{fig:spectra_mem_error} in the Appendix. Memory error arises from coherent phase accumulation and decoherence of qubits during idle periods between gate operations and can limit performance in trapped-ion quantum charge-coupled device (QCCD) architectures~\cite{PhysRevX.13.041052,DD_quantinuum_2026,Ransford2026}. To mitigate these errors and improve the accuracy of the spectrum, we found Pauli twirling and dynamical decoupling (DD) to be effective.

We also considered the effects of leakage errors, which occur when a qubit transitions to a state outside the computational subspace during circuit execution. Subsequent gate operations no longer act as intended and measurement of a leaked qubit generally yields the $\ket{1}$ state. Previous experiments on Quantinuum H1-1 showed that leakage error was significant when running deep, sequential circuits~\cite{nutzel25}. Prior to the hardware experiment, this behaviour was confirmed by executing one of the deepest circuits with a two-qubit gate depth of 1442 on the H2-1 device and analyzing the number of leaked shots.

These three techniques address distinct error channels: Pauli twirling suppresses the statistical effect of coherent over-rotations to a stochastic form on average, DD reduces idle-time memory errors, and leakage detection allows for the post-selection of erroneous shots in which the computation has left the computational subspace. Aside from leakage error post-selection which has a modest shot overhead, the use of Pauli twirling and DD incurs no additional shot overhead, unlike error mitigation approaches such as zero-noise extrapolation or probabilistic error cancellation.

\subsubsection{Pauli twirling}

The robustness of the Trotterized time evolution on hardware was improved by applying Pauli twirling, implemented in the form of echo compiling, to each Trotter step in the circuit. For a Pauli operator \(\hat P\in\{\hat I,\hat X,\hat Y,\hat Z\}\), we define the Pauli-conjugated Trotter step
\(
\hat P\,e^{-i\Delta t\,\hat P \hat H \hat P}\,\hat P .
\)
Because \(\hat P^2=\hat I\), all circuits realize the same ideal evolution \(e^{-i\Delta t \hat H}\).

In our implementation, the Trotter steps are deterministically cycled through the four variants
\begin{equation}
\hat I\,e^{-i\Delta t \hat H}\,\hat I,\qquad
\hat X\,e^{-i\Delta t\,\hat X\hat H\hat X}\,\hat X,\qquad
\hat Y\,e^{-i\Delta t\,\hat Y\hat H\hat Y}\,\hat Y,\qquad
\hat Z\,e^{-i\Delta t\,\hat Z\hat H\hat Z}\,\hat Z,
\end{equation}
and the sequence is then repeated. Thus, the first Trotter step is left unchanged, while subsequent steps are echoed with \(\hat X\), \(\hat Y\), and \(\hat Z\), respectively. The symmetrization is performed at the level of full Trotter steps rather than individual gates. Compared with gate-level Pauli twirling, this substantially reduces the number of distinct circuit instances that must be generated while still suppressing coherent errors.

\subsubsection{Dynamical decoupling}

We used the server-side DD scheme available on Quantinuum H2-1~\cite{DD_quantinuum_2026}. In this implementation, additional control pulses are inserted into idle regions of the circuit that exceed a user-specified threshold time. These pulses attempt to cancel accumulated phase errors by effectively reversing the sign of coherent evolution, thereby suppressing errors arising from slow variations in qubit frequencies, magnetic fields, and calibration imperfections. We used the default threshold time of $0.03$~s, i.e., the hardware compiler inserted additional DD pulses on qubits with idle times exceeding this threshold.

\subsubsection{Leakage error detection}

Leakage errors can be detected using a simple leakage-detection gadget~\cite{PhysRevX.13.041052,Stricker2020}, which requires one ancilla qubit prepared in the $\ket{1}$ state per system qubit; see Fig.~\ref{fig:gadget} in the Appendix for a circuit diagram. The gadget allows for post-selection by discarding shots in which leakage is detected via measurement of the ancilla register. For our hardware experiment, a leakage detection gadget was added to each system qubit.

\subsection{Circuit execution summary}

The hardware experiment used 21 system qubits (corresponding to 21 effective spin-\(\tfrac12\) degrees of freedom) and an additional 21 ancilla qubits for the leakage-detection gadgets. In total, 1420 circuits were executed on Quantinuum H2-1, accounting for 70 Trotter steps plus an initial $t=0$ step for state preparation, 10 different initial states, and $X$- and $Y$-basis measurements, i.e., $71 \times 10 \times 2 = 1420$ circuits. A total of $35\,500$ shots were used for the experiment, corresponding to 500 shots per Trotter step before leakage post-selection. Circuits were run in batches on the H2-1 device between January 21 and March 7, 2026; the device performance at the time of execution is listed in Table~\ref{tab:h2_params} in the Appendix. The average number of gates in each circuit per Trotter step is shown in Fig.~\ref{fig:num_gates} in the Appendix.

\section{Results}
\begin{figure}
\centering
\includegraphics[width=0.49\textwidth]{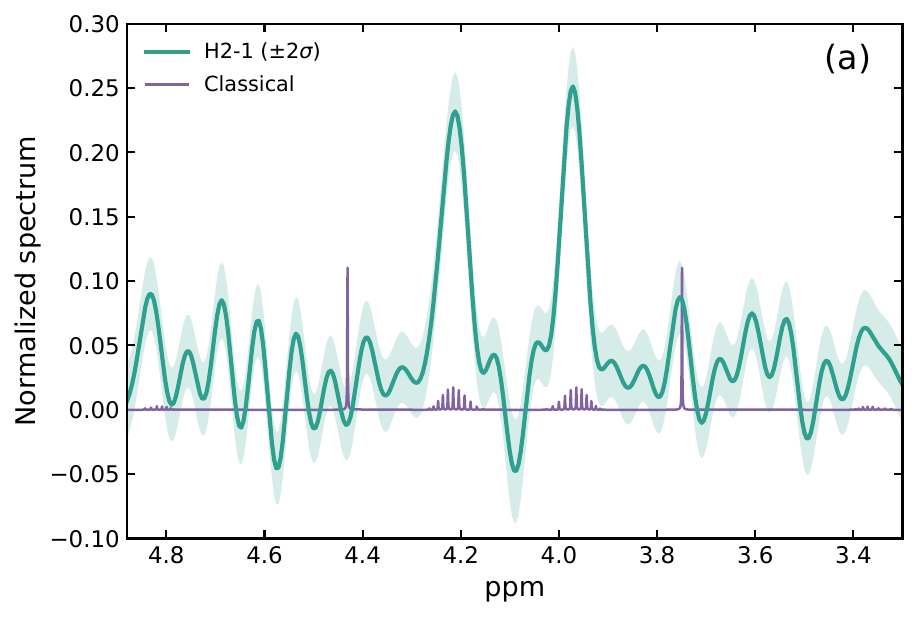}
\includegraphics[width=0.49\textwidth]{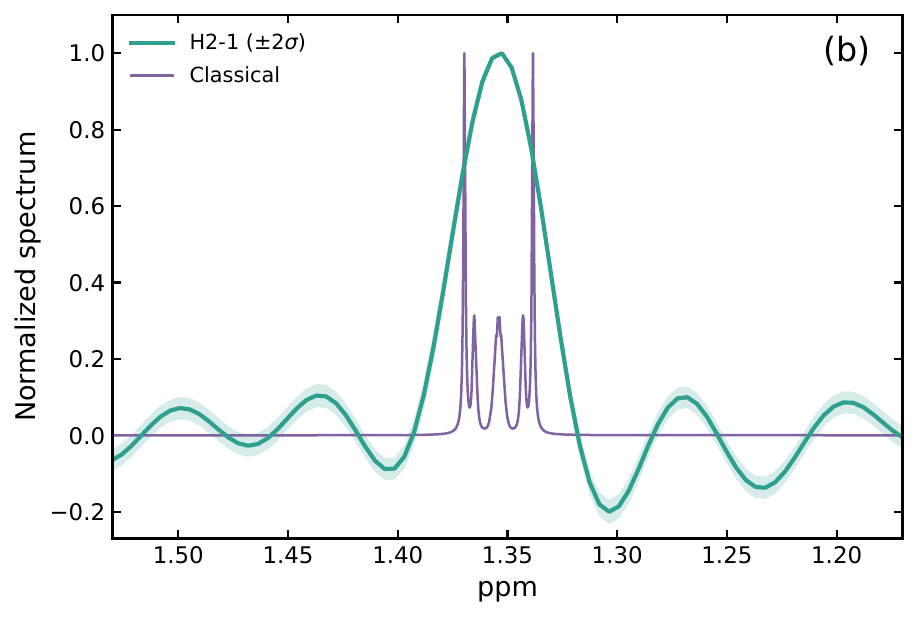}\\
\includegraphics[width=0.49\textwidth]{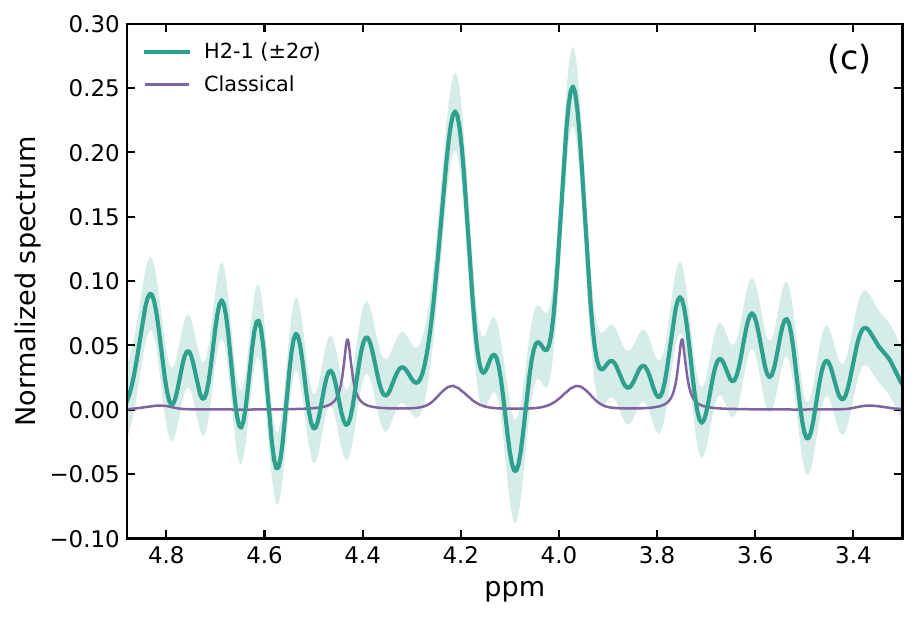}
\includegraphics[width=0.49\textwidth]{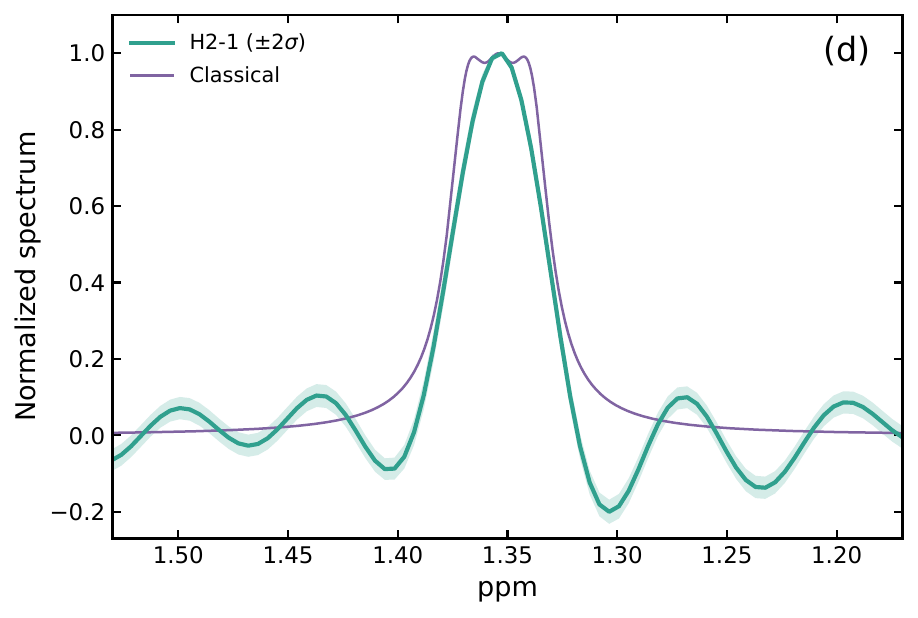}
\caption{Comparison of diphosphane NMR spectra obtained from simulations on the Quantinuum H2-1 trapped-ion quantum computer and from classical reference calculations. The H2-1 results are based on the projected Hamiltonian used in the hardware implementation, whereas the classical reference spectrum was calculated using the full 22-spin diphosphane Hamiltonian. Panels (a) and (b) show the classical spectrum with small line broadening, resolving the fine spectral structure, while panels (c) and (d) use larger line broadening to match the resolution of the quantum-hardware simulation~\cite{hqs_usecase,hqs_nmr,hqs_spectrum_tools}. The H2-1 hardware spectrum has been plotted with a pointwise bootstrap uncertainty band of width \(\pm 2 \sigma\) to illustrate the sensitivity of the spectrum to finite-shot sampling noise.}
\label{fig:hardware_spectrum}
\end{figure}

\begin{figure}
\centering
\includegraphics[width=0.49\textwidth]{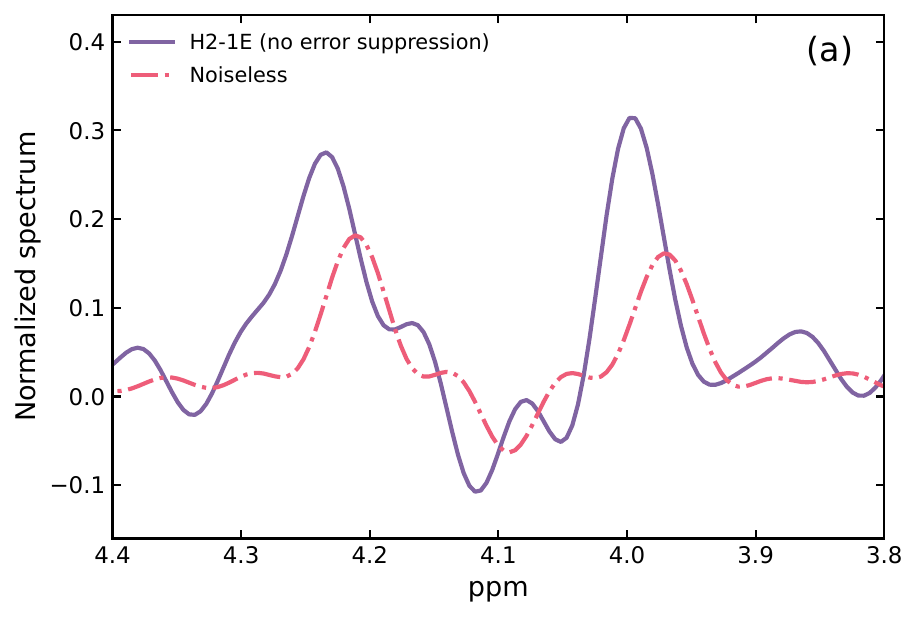}
\includegraphics[width=0.49\textwidth]{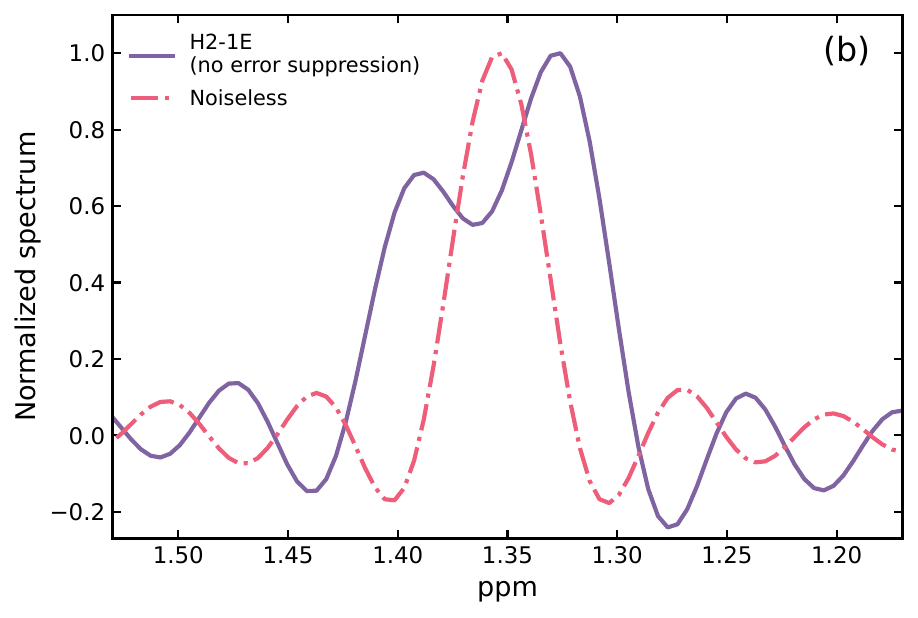}\\
\includegraphics[width=0.49\textwidth]{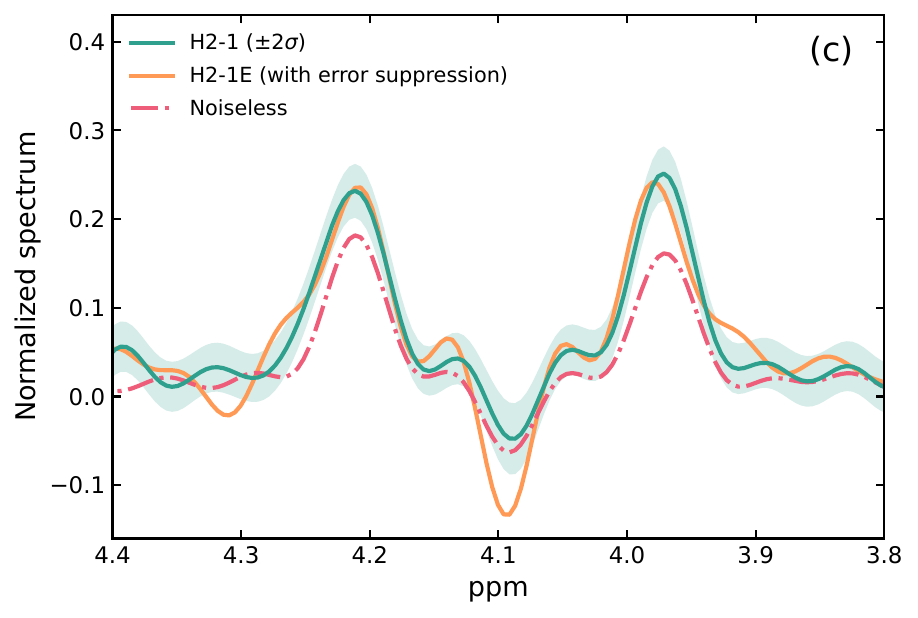}
\includegraphics[width=0.49\textwidth]{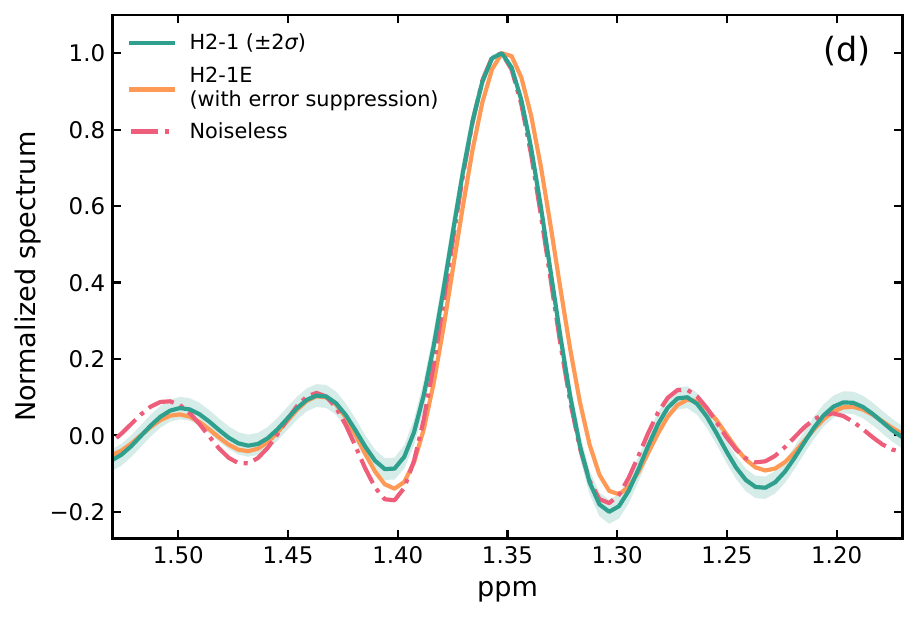}
\caption{Panels (a) and (b) compare diphosphane NMR spectra from noisy H2-1E emulator and noiseless simulations of 21-qubit circuits without error suppression. Panels (c) and (d) compare spectra from the H2-1 hardware, error-suppressed H2-1E emulator, and noiseless simulations. All emulator simulations used the same number of Trotter steps, shots per circuit, and total shot budget as the hardware experiment.}
\label{fig:hardware_emulator}
\end{figure}

\begin{figure}
\centering
\includegraphics[width=0.6\textwidth]{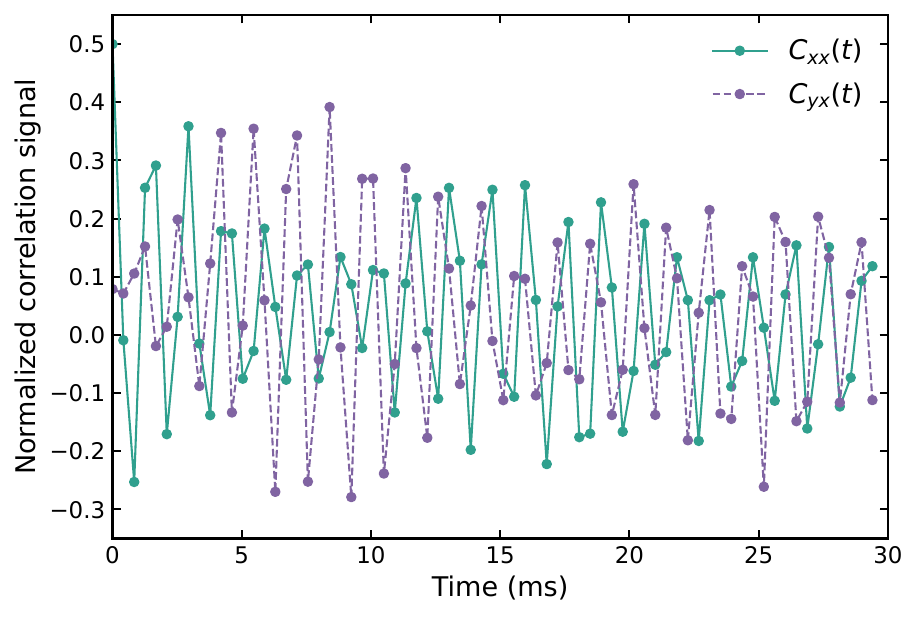}
\caption{Reconstructed time-domain NMR signal obtained from measurements on the Quantinuum H2-1 quantum computer, prior to the Fourier transform. The two plotted quadratures correspond to the correlation functions $C_{xx}(t)$ and $C_{yx}(t)$, which are combined as $s(t)\propto C_{xx}(t)+i\,C_{yx}(t)$ to form the complex free-induction decay (FID).}
\label{fig:time_signal}
\end{figure}

The NMR spectrum of diphosphane obtained from simulations on the Quantinuum H2-1 trapped-ion quantum computer is presented in Fig.~\ref{fig:hardware_spectrum} and Fig.~\ref{fig:hardware_emulator}. To illustrate the sensitivity of the spectrum to finite-shot sampling noise, the H2-1 hardware spectrum has been plotted with a pointwise bootstrap uncertainty band of width \(\pm 2 \sigma\). This was obtained from 2000 bootstrap resamples of the circuit-level measurement counts, with each resample propagated through the full spectrum-reconstruction pipeline. We compare against classical reference spectra computed using symmetry-based reductions~\cite{hqs_usecase,hqs_nmr}, as well as emulator simulations with and without noise and error suppression. All simulated spectra have normalized intensity and are plotted on the chemical-shift scale in ppm. Zero-padding was applied to the FID prior to the Fourier transform to smooth the displayed spectral profile. The correlation functions that define the FID, obtained from measurements on the H2-1 device, are shown in Fig.~\ref{fig:time_signal}.

The comparison with the classical calculations in Fig.~\ref{fig:hardware_spectrum} highlights two limitations of the quantum-hardware simulations. First, we were able to simulate only \(70\) Trotter steps, resulting in a maximum sampled FID duration of \(29.4~\mathrm{ms}\). This limits the achievable frequency resolution and introduces finite-time truncation artifacts, which appear in the spectrum as broadened peaks and oscillatory ripples. For a \(500\) MHz proton spectrometer, the corresponding spectral resolution is approximately \(0.07~\mathrm{ppm}\), which is too large to resolve the \(\sim 0.01~\mathrm{ppm}\) fine structure of interest, visible in panels (a) and (b) of the classical reference spectrum in Fig.~\ref{fig:hardware_spectrum}. Resolving these splittings would require a simulation time of at least \(T_{\max} \sim 200~\mathrm{ms}\). At the chosen timestep this would require over \(500\) Trotter steps, leading to circuit depths well beyond what is feasible on present-day hardware, even with the use of circuit gadgets to reduce the depth of NMR simulations~\cite{Burov2026}.

Secondly, the H2-1 simulations are based on the projected Hamiltonian adapted for the hardware implementation, whereas the classical reference spectrum was calculated using the full 22-spin diphosphane Hamiltonian. As a result, the H2-1 spectrum is missing two sharp peaks in the proton region from 3.3 to 4.9 ppm, which originate from spectral contributions that were removed by the projection of the phosphorus-spin pair.

Despite these limitations, the H2-1 hardware spectrum captures the main features of the classical reference spectrum, see panels (c) and (d) of Fig.~\ref{fig:hardware_spectrum} which shows a limited-resolution classical result. The dominant peak near 1.35~ppm in the methyl-proton region is reproduced, along with the distinctive double-peak structure near 4.1~ppm. Reproducing this weaker double-peak structure is significant, as it was not observed in previous quantum hardware demonstrations~\cite{Burov2025} or in simple classical clustering approaches~\cite{hqs_nmr}. Importantly, this agreement is not simply a consequence of line broadening or the absence of resolved fine splittings. Benchmarking against highly optimized classical NMR software packages such as {\it Spinach}~\cite{spinach2011} indicates that reproducing the double-peak structure, even without resolving the full multiplet fine structure, requires coupled-spin correlation effects up to fourth order, which is a nontrivial computational task~\cite{Kuprov_private}. With longer simulation times, we would expect the intensity of the double-peak structure in the H2-1 hardware spectrum to decrease and approach the magnitude seen in the classical reference spectrum, and for the fine structure to be resolved.

Additional simulations were performed using the Quantinuum H2-1E emulator with the same number of Trotter steps, shots per circuit, and total shot budget of the hardware experiment. To simulate the 42-qubit circuits executed on hardware, we implemented the leakage detection gadgets with 7 ancilla qubits instead of 21 ancilla qubits, enabling 28-qubit statevector simulations. Simulations of 21-qubit circuits with no error suppression applied, i.e., Trotterized time evolution of the 21 effective-spin Hamiltonian without Pauli twirling, dynamical decoupling, or leakage detection, were also carried out using the H2-1E emulator, and an ideal noiseless backend.

Panels (a) and (b) of Fig.~\ref{fig:hardware_emulator} show the spectra obtained from simulations of the 21-qubit circuits without error suppression using the noisy H2-1E emulator and a noiseless backend. The double-peak structure in the $3.9$--$4.3$~ppm region is noticeably shifted due to the effects of device noise, while the dominant peak at $\approx 1.35$~ppm exhibits an incorrect broadened structure compared to the noiseless spectrum. Panels (c) and (d) of Fig.~\ref{fig:hardware_emulator} show the H2-1 hardware spectrum, the H2-1E emulator spectrum obtained from the 28-qubit circuits with error suppression, and noiseless results. The close agreement between the H2-1 hardware and H2-1E emulator spectra validates the accuracy of the noise model of the emulator. Furthermore, the close agreement with the noiseless spectrum highlights the effectiveness of the error suppression techniques at combatting device noise and preserving the relevant spectroscopic structure from simulations on the quantum device.

\begin{figure}
\centering
\includegraphics[width=0.49\textwidth]{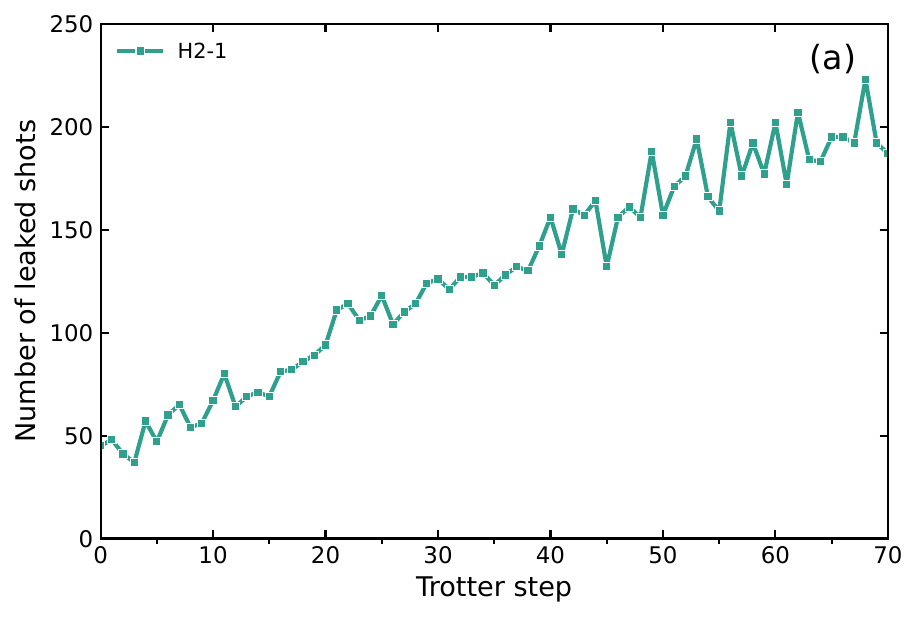}
\includegraphics[width=0.49\textwidth]{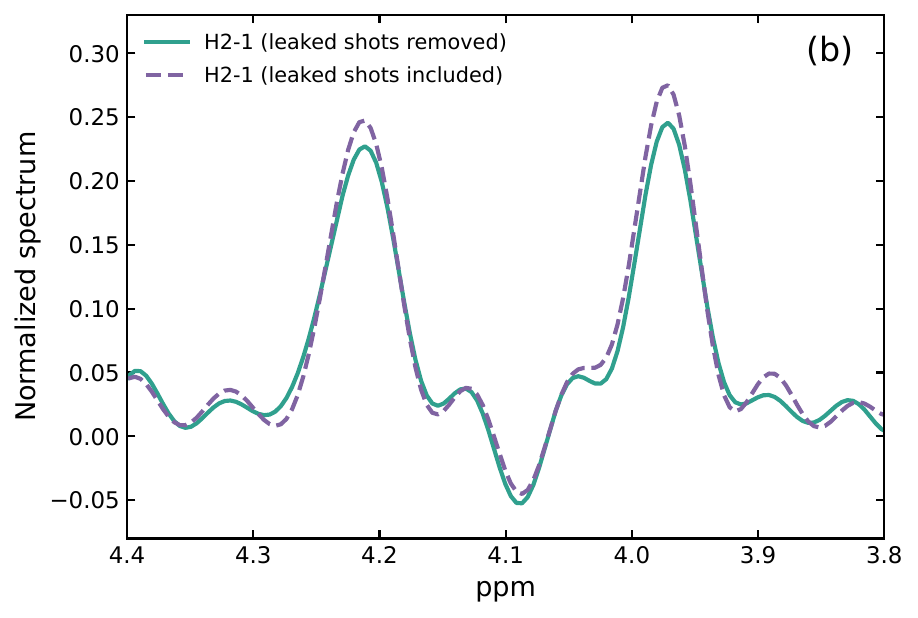}
\caption{(a) Total number of leaked shots per Trotter step for NMR simulations on the Quantinuum H2-1 trapped-ion quantum computer. Each Trotter step used 500 shots across 20 circuits. (b) Simulated spectrum obtained from the H2-1 device with and without leaked shots included in the FID dataset before the Fourier transform.}
\label{fig:leakage_trotter}
\end{figure}

Although memory error was the dominant error source in simulations, see Fig.~\ref{fig:spectra_mem_error}, prior to the hardware experiment it was unclear how significant leakage errors would be. The impact of leakage error on the experiment is shown in Fig.~\ref{fig:leakage_trotter}. In panel (a), the number of leaked shots per Trotter step is illustrated for simulations on the H2-1 device. As expected, leakage errors increase with the number of Trotter steps and increasing circuit depth. Our experiment used a total of $35\,500$ shots, of which $25.7\%$ of shots exhibited leakage with a larger proportion leaked for deeper circuits. This observation is consistent with a previous experiment on Quantinuum H1-1, which found leakage to be higher when executing deep, sequential circuits~\cite{nutzel25}.

In panel (b) of Fig.~\ref{fig:leakage_trotter}, the spectrum in the $3.9$--$4.3$~ppm region is reconstructed from the H2-1 hardware experiment with leaked shots included in the FID dataset, and compared to the final H2-1 spectrum where leaked shots had been removed. Post-selection only marginally changes the peak intensities while the peak positions are almost unchanged within the resolution of the hardware experiment. These results indicate that leakage errors can be managed sufficiently to preserve the relevant spectral features.

\section{Outlook}

In this work, we have executed an end-to-end digital NMR simulation workflow on present-day trapped-ion quantum hardware at a scale and fidelity sufficient to recover nontrivial molecular spectral structure. This was done in a scalable manner using a Trotterized implementation of time evolution, together with controlled Hamiltonian approximations and targeted error suppression. The use of Pauli twirling, DD, and leakage detection effectively suppressed hardware noise, most notably memory error, which was the dominant source of error in simulations on Quantinuum H2-1. These suppression techniques do not directly correct stochastic Pauli or depolarizing errors, yet still produced a spectrum in close agreement with noiseless emulator results, suggesting that NMR simulation may be somewhat robust to hardware noise. Furthermore, apart from the modest reduction in usable shots as a result of leakage post-selection, Pauli twirling and DD incurred no additional shot overhead.

Our findings strengthen the case for NMR simulation as a promising and industrially-relevant application area for near- to mid-term quantum advantage. Achieving this, however, requires outperforming highly optimized classical solvers that exploit problem-specific structure to tackle increasingly large and complex systems. In the case of high-field liquid-state NMR, classical methods appear to cover a large portion of practically relevant use cases, and identifying molecular systems beyond their reach remains an open challenge. A more promising regime is zero- to ultralow-field (ZULF) NMR, where internal spin-spin couplings dominate over Zeeman interactions and cannot be neglected~\cite{Barskiy2025}, leading to dynamics that are harder to simulate classically. Recently, a rigorous resource estimation analysis using fault-tolerant quantum algorithms, specifically qubitized quantum dynamics, was applied to liquid-state ZULF NMR simulations~\cite{Elenewski2026}. Logical qubit and {\it T}-gate counts were evaluated for thousands of small organic molecules and biological macromolecules, suggesting that NMR simulation on quantum computers could have practical utility in the fault-tolerant era.

As quantum hardware continues to advance, it is important to continue developing large-scale hardware experiments to assess the practical implementation of quantum algorithms, as their sensitivity to different error sources can vary substantially. Hamiltonian simulation is one of the leading candidates for quantum advantage and a broad range of algorithms have been developed for both near-term and fault-tolerant applications. Understanding how these algorithms perform in the presence of device noise, under different compilation schemes, and within resource constraints will be essential for translating asymptotic advantage into useful applications. For example, recent work has shown that Trotterization can be more resource-efficient than quantum signal processing or qubitization-based approaches for Hamiltonian simulation, depending on the problem structure and compilation scheme~\cite{LeBlond2025,Blunt2025}.

\section{Acknowledgements}
The authors thank Ilya Kuprov for valuable discussions, and Maud Einhorn, Joshua Savory, and Matthew Girling for assistance with the collaboration and hardware experiment. The authors thank Matthew Girling, Georgia Prokopiou, and Charles Baldwin for comments on the manuscript.

\section{Author contributions}
PS and MM developed the quantum NMR simulation methodology. AO performed the quantum hardware and emulator simulations with input from PS, EG and DMR. All authors conceived the study, analyzed the results, and contributed to writing the manuscript.

\bibliographystyle{apsrev4-1}
\bibliography{references.bib}

\newpage

\appendix
\phantomsection
\section*{Appendix}
\label{sec:appendix}

\renewcommand{\thefigure}{A.\arabic{figure}}
\setcounter{figure}{0}

\renewcommand{\thetable}{A.\arabic{table}}
\setcounter{table}{0}

\begin{figure}[h]
\centering
\includegraphics[width=\textwidth]{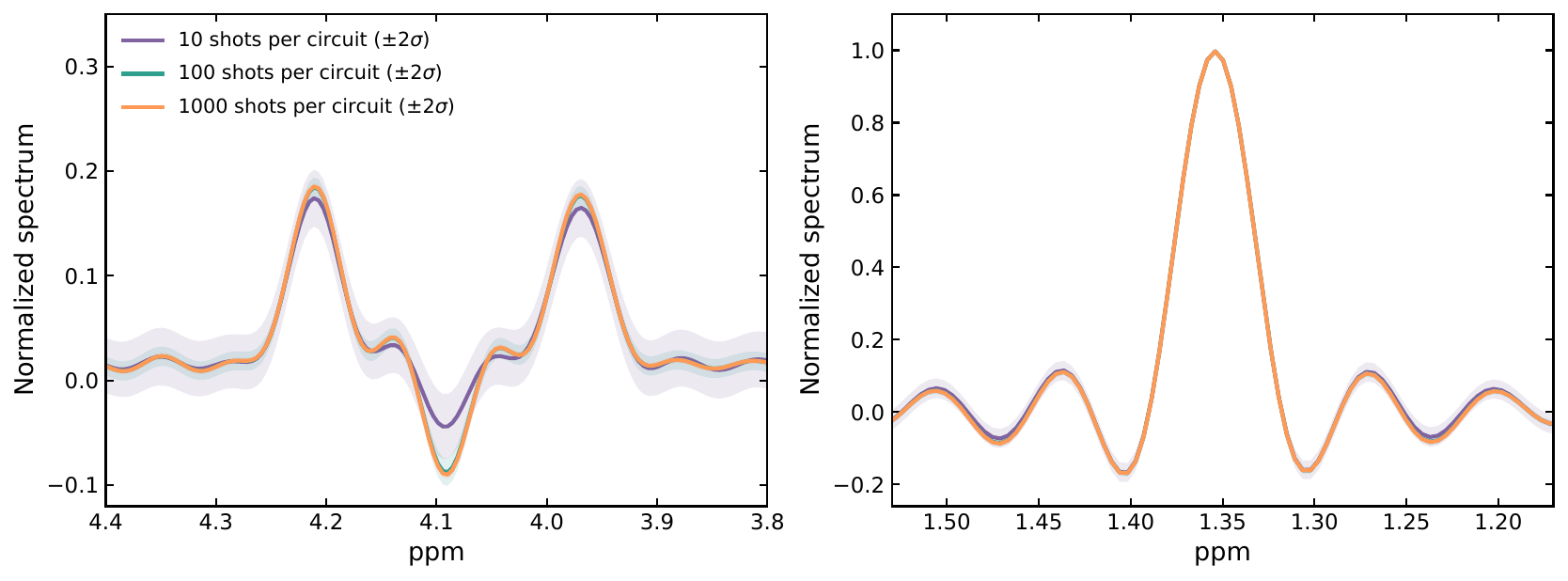}
\caption{Simulated NMR spectra showing the impact of shot noise. The noiseless simulations are of 21-qubit circuits without error suppression for 10, 100, and 1000 shots per circuit. Each spectrum has been plotted with a pointwise bootstrap uncertainty band of width \(\pm 2 \sigma\) (obtained from 2000 bootstrap resamples) to illustrate the sensitivity of the spectrum to finite-shot sampling noise.}
\label{fig:spectra_shot_noise}
\end{figure}

\begin{figure}[h]
\centering
\includegraphics[width=0.49\textwidth]{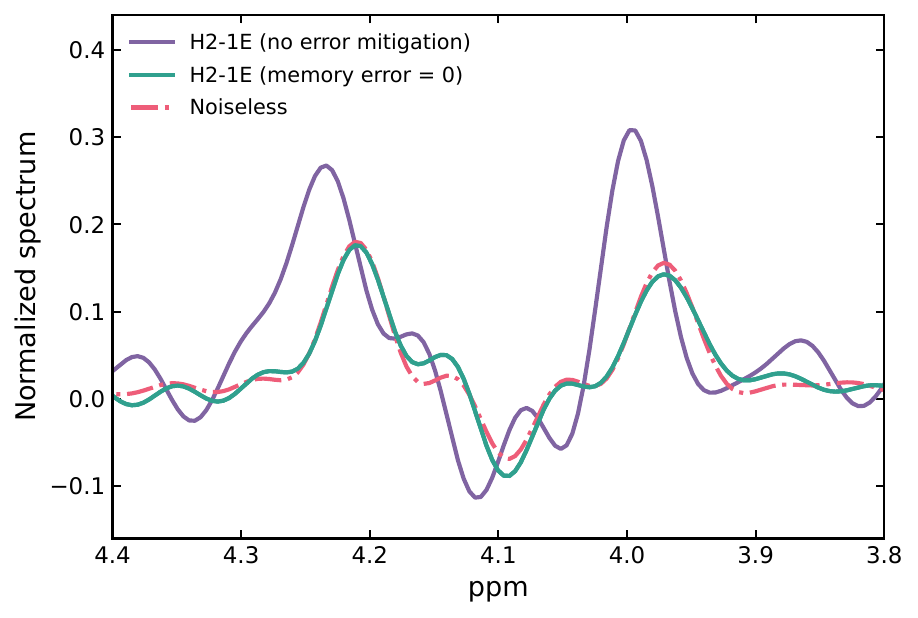}
\includegraphics[width=0.49\textwidth]{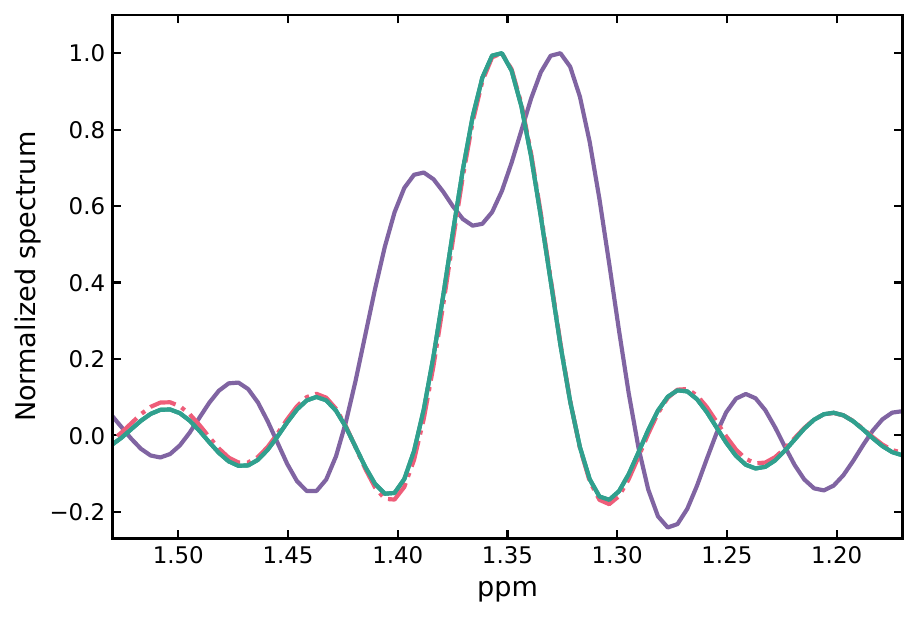}
\caption{Simulated NMR spectra showing the impact of memory error. The H2-1E emulator and noiseless simulations are of 21-qubit circuits without error suppression. When memory error was disabled in the H2-1E emulator noise model, the simulated spectrum showed close agreement with the ideal noiseless result.}
\label{fig:spectra_mem_error}
\end{figure}

\begin{figure}[h]
\centering
\includegraphics[width=0.8\textwidth]{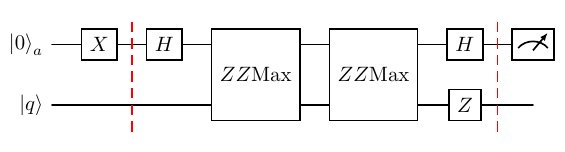}
\caption{Leakage detection gadget~\cite{PhysRevX.13.041052}. An ancilla qubit $\ket{0}_a$ is used to detect whether qubit $\ket{q}$ has leaked. The ancilla is prepared in $\ket{1}$ by the $\hat{X}$ operation before application of the gadget, enclosed by the red dashed lines. If $\ket{q}$ has leaked, the two-qubit $ZZ{\rm{Max}}=\exp(-i\frac{\pi}{4}(\hat{Z}\otimes\hat{Z}))$ gates have no logical effect and the ancilla is measured as $\ket{1}$. Otherwise, if $\ket{q}$ has not leaked, the gadget acts as $\hat{X}_a\otimes \hat{I}_ q$ and the ancilla is measured as $\ket{0}$. This enables post-selection of measurement shots based on the state of the ancilla qubit register.}
\label{fig:gadget}
\end{figure}

\begin{table}[h]
\centering
\begin{tabular}{l @{\hspace{1cm}} c}
\hline\hline
\textbf{Error Type} & \textbf{Infidelity}\\
\hline
1-qubit gate error & $1.9 \times 10^{-5}$\\
2-qubit gate error & $1.1 \times 10^{-3}$ \\
1-qubit leakage error & $4.8 \times 10^{-6}$ \\
2-qubit leakage error & $1.4 \times 10^{-4}$ \\
Memory error & $2.0 \times 10^{-4}$ \\
Measurement crosstalk error & $6.6 \times 10^{-6}$ \\
SPAM error & $1.4 \times 10^{-3}$ \\
\hline\hline
\end{tabular}
\caption{Hardware specification of the Quantinuum H2-1 trapped-ion quantum computer at the time of the experiment (between January 21 and March 7, 2026). SPAM error refers to the state preparation and measurement error for the $\ket{1}$ state, which is larger than the error for the $\ket{0}$ state.}
\label{tab:h2_params}
\end{table}

\begin{figure}[h]
\centering
\includegraphics[width=0.6\textwidth]{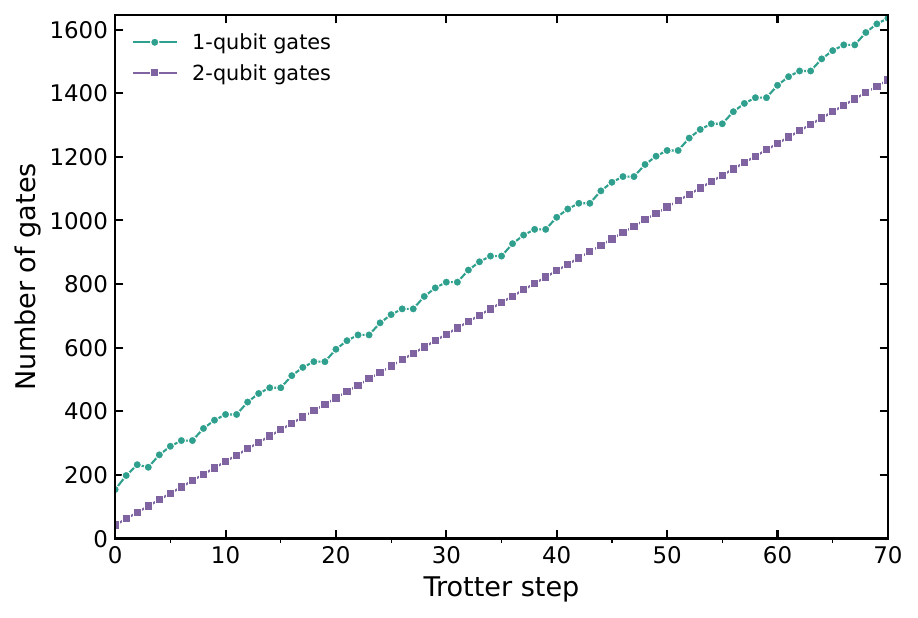}
\caption{The average number of one- and two-qubit gates per Trotter step for the circuits executed in the Quantinuum H2-1 hardware experiment.}
\label{fig:num_gates}
\end{figure}

\end{document}